\documentclass[prl,10pt,twocolumn,superscriptaddress,showpacs]{revtex4}
\usepackage{amsmath}
\usepackage{latexsym}
\usepackage{amssymb}
\usepackage{graphics,epstopdf}
\usepackage{graphicx}
\usepackage[colorlinks=true, citecolor=blue, urlcolor=blue ]{hyperref}
\usepackage{float}
\usepackage{graphicx}
\usepackage{amsfonts}
\begin{document}

\title{ Optimal violation of Leggett-Garg inequality for arbitrary spin and emergence of classicality through unsharp measurement }

\author{Shiladitya Mal}
\email{shiladitya.27@gmail.com}

\author{A. S Majumdar}
\email{archan@bose.res.in}
\affiliation{S. N. Bose National Centre for Basic Sciences, Block JD, Sector III, Salt Lake, Kolkata  700098, India}

\begin{abstract}
We consider temporal correlations of particles with arbitrary spin. We show
that the Leggett-Garg inequality can be maximally violated irrespective of the
value of spin, thus improving upon an earlier result by Kofler and Brukner
[Phys. Rev. Lett. \textbf{99}, 180403 (2007)].  Our proof is accomplished 
through a suitable adoption
of a measurement scheme which has previously been employed for studying the 
spatial correlation in a system with arbitrary spin. We next consider 
generalized or unsharp measurements as a method for coarse graining in a 
quantitative manner, and show that this inequality can not be violated below a precise
value of the sharpness parameter. We then apply the Fine's theorem in
context of the Leggett-Garg (LG) inequality in order to derive LG-CH
inequalities which provide a sufficient condition for emergence of
classicality.
\end{abstract}


\maketitle

\section{\textbf{I. Introduction}}

Quantum physics has been enormously successful in 
explaining nonclassical manifestations of nature. Bell's theorem~\citep{Bell} 
proves through quantum violation of the Bell-CHSH~\citep{CHSH} inequality, that 
correlations arising from spatially separated quantum systems cannot be 
explained if we adopt local realism- a genuine classical world view.  
Quantum mechanical violation of the Bell-CHSH inequality for 
dichotomic measurements is upper bounded by $2\sqrt{2}$~\citep{tsi}. 
The question as to how quantum correlations are fundamentally limited has been 
explored as an optimization problem which can be done efficiently, and assured 
to reach the global optimum since it represents a so-called semidefinite 
program~\citep{vanden}.  Upper bounds on the quantum violation 
of more general Bell-type inequalities have also been obtained~\citep{wehner}. 

On the other hand, from the very beginning of quantum mechanics it has been 
of great interest
to understand how the quanutum-classical transition occurs. The classical 
limit problem arises if quantum mechanics is taken to be a universal theory
encompassing classical physics. One might then expect the 
typically classical behavior of a macrosystem to emerge within a quantum 
mechanical description if the relevant quantities are large compared to 
the Planck's constant. However, it has been shown that quantum features persist 
in the limit of large quantum numbers such as the number of constituents of
a system~\citep{mer1}, or the value of its spin angular momentum~\citep{M,P,GP}.
In such cases the magnitude of violation of the relevant local realist 
inequalities generally seems to increase even in the limit of large numbers
of particles and large spins considered simultaneously~\citep{HM}.
Nonclassicality of multiparty and multilevel systems in the context of
 spatial correlation have been extensively studied, belying expectations of 
classical properties emerging automatically for `large' quantum systems~\citep{GM,how}. 

Besides the issue of local realism, Leggett and Garg in a seminal 
paper~\citep{lgi1} have provided a way to test quantum mechanics in the 
macroscopic domain by deriving an inequality based on macroscopic realism- a 
classical world view dealing with temporal correlations. The
two assumptions of macrorealism(MR) are, (i) {\it macrorealism per se:} a 
macroscopic object which has available to it two or more macroscopically 
distinct states is at any given time in a definite one of those states, and
(ii) {\it noninvasive measurability (NIM):} it is possible in principle to 
determine which of these states the system is in without any effect on the 
state itself or on the subsequent system dynamics. Another assumption 
implicit in the theory is induction which states that system properties 
are not determined by the final conditions. Various studies 
have been performed recently on theoretical aspects of the Leggett-Garg 
inequality~(LGI) and its generalizations~\citep{lgi2,KB,avis,KB2,budroni}, along 
with a couple insightful
reviews~\citep{revw}. Several experiments testing the violation of LGI have 
also been reported~\citep{ex1}.
It has been of considerable interest to investigate the extent to which various 
Bell ienqualities are violated by quantum mechanics for different types of 
systems~\citep{onof}.  

Within the context of temporal correlations, the question as to how the 
classical world emerges out of quantum physics has been discussed  by Kofler 
and Brukner~\citep{KB}. Inspired by the earlier ideas of Peres~\citep{peres1} 
on the classical limit of quantum mechanics, they have presented a different 
theoretical approach to macroscopic realism and classical physics within 
quantum theory. They showed that if consecutive eigenvalues of the spin 
component can be sufficiently resolved, the LGI will be violated for arbitrary 
large spin. On the other hand, with sufficiently coarse grained measurement, 
classical laws would emerge for a macroscopic system with very large dimension.
This approach is rather different from the decoherence program. 
However, the violation they obtained for large spin is not maximal. It remains
unclear as to why the violation is lesser that the value $2\sqrt{2}$ achieved 
for spin $1/2$ particles and remains constant asymptotically for large spin.
 The choice of observables may indeed have a role to play in
the quantum violation of the LGI. 

  In 
order to obtain optimal violation of the LGI for arbitrary 
spin, we employ a variant of the measurement scheme suggested earlier
by Gisin and Peres~\citep{GP} in the context of testing local realism. 
Interestingly, for the case of spatial correlations the above measurement
 scheme~\citep{GP} yields maximal violation of a local realist inequality
only for half-integral spin 
systems. For integral spin systems the amount of violation drops, and the value 
of $2\sqrt{2}$ is achieved only when the spin becomes infinitely large. 
It has earlier 
been shown~\citep{fritz} that the success probability for Hardy type
protocols~\cite{hardy} is much higher in temporal correlations than in spatial
ones. Our finding is thus indicative of another aspect of difference between 
spatial and temporal correlations in addition to the main results.

In the second part of our work we focus on the issue of quantum-classical 
transition by adopting unsharp measurement as a completely different method of 
coarse graining from the method followed in \citep{KB}.  Unsharp measurement, a form 
of positive operator valued measurement (POVM), is well studied in 
the quantum formalism~\citep{bookB,BJ}.
 Emergence of classicality is argued by transforming the LGI to a necessary 
and sufficient condition for macrorealism for dichotomic measurement 
at four different time following Fine's theorem~\citep{fine}.
We find that  below a definite degree of sharpness of measurement, a 
macrorealist inequality like LGI can not be violated, and hence, there exists 
a non-invasive realist model compatible with classical physics for reproducing 
correlation statistics. Through this formalism we find a way to compare 
between different inequalities based on different assumptions, as also 
discussed in a recent work~\citep{wlgi} involving one of us. 

The main results obtained in this work are: (i) we obtain violation of LGI up to
 $2\sqrt{2}$ for arbitrary spin which is optimal for dichotomic measurements, (ii) by
application of unsharp measurement we show that satisfaction of the LGI ensues
below a specific value of the sharpness parameter, and (iii) application of the 
Fine's theorem in context of the LGI leads to LG-CH type of inequalities which
provide a sufficient condition for MR.
The paper is organized as follows. In the next section we briefly discuss the
derivation of the LGI based on an ontological model~\citep{years}, and 
provide a quick review of its quantum violation for arbitrary value of spin using the 
formalism of Ref.~\citep{KB}. In Section III we employ another scheme of 
measurement~\citep{GP} to show that optimal violation for arbitrary spin can 
be obtained. We further describe a process of conceptual experimental 
realization of the scheme. In Section  IV  we apply the formalism of unsharp 
measurement on the  LGI, presenting a different method of coarse graining 
responsible for the quantum-to-classical transition. In Section V we apply Fine's theorem to 
temporal correlation and show how  LGI becomes a  
sufficient condition for MR. Some concluding remarks are presented in Section V.

\section{\textbf{II. The Leggett-Garg inequality}}

 Following the ontological framework discussed in\citep{years}, we 
begin with a short derivation of the LGI. In this framework any Heisenberg 
picture operator in quantum mechanics can be written as an average over a set 
of hidden variables ${\lambda}$. The role of the initial state is to provide a 
probability distribution on the set of hidden variables, which we denote as 
$\rho(\lambda)$, called the ontic state. The average of an observable can be 
written as
\begin{eqnarray}
<\hat{A}(t)> =\int d\lambda A(\lambda, t) \rho(\lambda),
\end{eqnarray}
where $A(\lambda, t)$ is the value taken by the observable on the hidden 
variable $\lambda$. The correlation between two observables is given by
\begin{eqnarray}
<\hat{B}(t_{2})\hat{A}(t_{1})>=\int d\lambda B(\lambda, t_{2})A(\lambda, t_{1})\rho(\lambda |A, t_{1}).
\end{eqnarray}
Non-invasive measurability (NIM) can be defined as $\rho(\lambda |A, t_{1}, B, t_{2}...)=\rho(\lambda)$, i.e., a measurement performed does not change the 
distribution of $\lambda$ (like the locality condition in Bell's theorem). Let 
us take $A, B$ as observables measured on a single system at different times 
denoted by $Q(t_{1}), Q(t_{2})$ . Now, following similar steps as in the 
derivation of the Bell inequality, one obtains
\begin{eqnarray}\label{der}
<\hat{Q}(t_{2})\hat{Q}(t_{1})> - <\hat{Q}(t_{4})\hat{Q}(t_{1})>  \nonumber\\
=\int d\lambda [Q(\lambda, t_{2})Q(\lambda, t_{1})- Q(\lambda, t_{4})Q(\lambda, t_{1})]\rho (\lambda |Q,t_{1}) \nonumber\\
=\int d\lambda Q(\lambda, t_{2})Q(\lambda, t_{1})[1\pm Q(\lambda, t_{4})Q(\lambda, t_{3})]\rho(\lambda |Q, t_{1}) \nonumber\\
- \int d\lambda Q(\lambda, t_{4})Q(\lambda, t_{1})[1\pm Q(\lambda, t_{3})Q(\lambda, t_{2})]\rho(\lambda |Q, t_{1}).
\end{eqnarray}
Now,
\begin{eqnarray}
 |<\hat{Q}(t_{2})\hat{Q}(t_{1})> - <\hat{Q}(t_{4})\hat{Q}(t_{1})> |
\leq 2\pm \\ \nonumber [\int d\lambda Q(\lambda ,t_{4})Q(\lambda , t_{3})\rho (\lambda |Q,t_{1})\\ \nonumber +\int d\lambda Q(\lambda ,t_{3})Q(\lambda , t_{2})\rho (\lambda |Q,t_{1}) ].
\end{eqnarray}
Invoking NIM, we have,
\begin{eqnarray}
|<\hat{Q}(t_{2})\hat{Q}(t_{1})> - <\hat{Q}(t_{4})\hat{Q}(t_{1})> | \mp \\ \nonumber
 [<\hat{Q}(t_{3})\hat{Q}(t_{2})> + <\hat{Q}(t_{4})\hat{Q}(t_{3})> ] \le 2.
\label{lgineq}
\end{eqnarray}
This is four term Leggett-Garg inequality.

In an actual experiment, $Q(t)$, a macrovariable measured at time $t$, is 
found to take a value $+1(-1)$ depending on whether the system is in the state 
$1(2)$. We consider series of measurements with the same initial conditions 
such that in the first series $Q$ is measured at times $ t_{1}$ and $t_{2}$, in 
the second at $t_{2}$ and $t_{3}$, in the third at $t_{3}$ and $t_{4}$, and in 
the fourth at $t_{1}$ and $t_{4}$ (here $ t_{1} < t_{2} < t_{3} < t_{4}$). 
From such measurements one obtains the temporal correlations $C_{ij}=\langle Q_{i}Q_{j}\rangle =p^{++}(Q_{i},Q_{j})-p^{-+}(Q_{i},Q_{j})-p^{+-}(Q_{i},Q_{j})+p^{--}(Q_{i},Q_{j})$, where $p^{++}(Q_{i},Q_{j})$ is the joint probability of getting `$+$' 
outcomes at both times $t_{i}$ and $t_{j}$. Experimentally, these joint 
probabilities are determined from the Bayes' rule $p^{++}(Q_{i},Q_{j})=p^{+}(Q_{i})p^{+|+}(Q_{j}|Q_{i})$, where $p^{+|+}(Q_{j}|Q_{i})$ is the conditional probability 
of getting `$+$' outcome at $t_{j}$ given that `$+$' outcome occurs
at $t_{i}$.

Let us now briefly describe how quantum violation of the LGI was obtained 
in\citep{KB}. Consider precession of a spin $1/2$ particle under 
the unitary evolution $U_{t}=e^{-i\omega t\sigma_{x}/2}$, where $\omega$ is the 
angular precession frequency. Measurement of $\sigma_{z}$ at times $t_{1}$ and 
$t_{2}$ yields the temporal correlation $C_{12}=\cos\omega(t_{2}-t_{1})$. Here 
the state transformation rule is given by $\rho\rightarrow P_{\pm}\rho P_{\pm}/Tr[P_{\pm}\rho P_{\pm}]$. Choosing equidistant measurement times with time 
difference $\Delta t=t_{2}-t_{1}=\pi/4\omega$, the maximum value taken by
the l.h.s of Eq.(\ref{lgineq}) is given by $2\sqrt{2}$. For a spin $j$ system 
with a maximally mixed initial state $\frac{1}{2j+1}\sum_{m=-j}^{m=+j}\vert m\rangle\langle m\vert$,  evolving unitarily under $U_{t}=e^{-i\omega t\hat{J_{x}}}$, 
measurement of the dichotomic parity operator $\sum_{m=-j}^{m=+j}(-1)^{j-m}\vert m\rangle\langle m\vert$, leads to the two-time correlation function given by
\begin{eqnarray}
 C_{12}=\sin[(2j+1)\omega\Delta t]/(2j+1)\sin[\omega\Delta t].
\end{eqnarray}
With these correlations the LGI expressed as 
$K=C_{12}+C_{23}+C_{34}-C_{14}\leq 2$ becomes
\begin{eqnarray}
K=\frac{3\sin x}{x}-\frac{\sin 3x}{3x}\leq 2,
\end{eqnarray}
where $x=(2j+1)\omega\Delta t$. For $x\approx 1.054$, the maximal violation 
in this case is 
obtained for infinitely large $j$, with the value $2.481$, i.e., $42$ percent
short of the largest violation of $2\sqrt{2}$ allowed by quantum theory. 

\section{\textbf{III. Optimal violation for arbitrary spin}}

We now show how the maximum correlation up to $2\sqrt{2}$ which is the upper 
bound of quantum theory for dichotomic measurements, can be achieved not only 
for spin $1/2$ particles, but for systems having arbitrary spin too.\\
{\it\textbf{Lemma}:} If a dichotomic observable $Q$ is measured successively 
at times $t_{i}$ and $t_{j}$ on any state 
$\rho$ of a two dimensional system evolving unitarily,  then the two-time 
correlation function is given by $C_{ij}=\frac{1}{2}tr[Q(t_{i})Q(t_{j})]$.
Here $Q(t_{i})=U^{\dagger}(t_{i})QU(t_{i})$ and $Q(t_{j})=U^{\dagger}(t_{j})QU(t_{j})$ are time evolved observables in the Heisenberg picture.\\
{\it\textbf{Proof}:} The initial state $\rho$ is evolved to $U(t_{i})\rho U^{\dagger}(t_{i})$. At $t_{i}$, $Q$ is measured and according to the outcome $`\pm$', 
the post-measurement state becomes $P_{\pm}U(t_{i})\rho U^{\dagger}(t_{i})P_{\pm}/tr[P_{\pm}U(t_{i})\rho U^{\dagger}(t_{i})P_{\pm}]$, where $P_{\pm}$ are the two 
orthogonal projectors of the observable $Q$, and $P_{\pm}U(t_{i})\rho U^{\dagger}(t_{i})P_{\pm}/p_{\pm}=P_{\pm}$. Here, $p_{\pm}=tr[P_{\pm}U(t_{i})\rho U^{\dagger}(t_{i})]$, are probability of getting outcomes `$\pm$'. Again, this post-measurement 
state is evolved to time $t_{j}$ and becomes $U(\Delta t)P_{\pm}U^{\dagger}(\Delta t)$, with $\Delta t=t_{j}-t_{i}$. Now, the conditional probabilities are given by $p_{k|l}=tr[P_{k}U(\Delta t)P_{l}U^{\dagger}(\Delta t)$, where $p_{k|l}$ denotes 
the probability of getting an outcome $k$ at time $t_{j}$ when the outcome $l$ 
occurs at time $t_{i}$. Hence, the two-time correlation is given by
\begin{eqnarray}
C_{ij}=p_{+}(p_{+|+}-p_{-|+})+p_{-}(p_{-|-}-p_{+|-})\nonumber\\
= p_{+}(tr[(P_{+}-P_{-})U(\Delta t)P_{+}U^{\dagger}(\Delta t)])\nonumber\\ 
+ p_{-} (tr[(P_{-}-P_{+})U(\Delta t)P_{-}U^{\dagger}(\Delta t)])\nonumber\\
\end{eqnarray}\\
Now, using $P_{+}-P_{-}=Q$, $tr[Q]=0$, $p_{+}+p_{-}=1$, and the cyclic property
of the trace, we have $C_{ij}=tr[P_{+}U^{\dagger}(\Delta t)Q U(\Delta t)]$.
Since, $P_{+}=\frac{\mathbb{I}+Q}{2}$ and $U(\Delta t)=U(t_{2})U^{\dagger}(t_{1})$, 
 we finally have
\begin{eqnarray}
C_{ij}=tr[QU^{\dagger}(\Delta t)Q U(\Delta t)]/2\nonumber\\
=tr[U^{\dagger}(t_{1})Q U(t_{1})  U^{\dagger}(t_{2})Q U(t_{2})]/2 .
\end{eqnarray}\\
This completes the proof of the lemma.\\ 
{\it \textbf{Theorem}:} For any state of a single quantum system of arbitrary 
spin  there exists observables with eigenvalues $\pm 1$ and a measurement 
scheme such that the Leggett-Garg inequality can be maximally violated.\\
{\it\textbf{ Proof}:} Let $\Gamma_{x}, \Gamma_{y}, \Gamma_{z}$ be block-diagonal 
matrices, in which each block is an ordinary Pauli matrix, $\sigma_{x}, \sigma_{y}$ and $\sigma_{z}$ respectively. The only nonvanishing elements are given by
\begin{eqnarray}
(\Gamma_{x})_{2n-1,2n}=(\Gamma_{x})_{2n,2n-1}=1\nonumber\\
(\Gamma_{y})_{2n-1,2n}=i,      (\Gamma_{y})_{2n,2n-1}=-i\nonumber\\
(\Gamma_{z})_{n,n}=(-1)^{n-1}
\end{eqnarray}\\
Suppose mixed states of spin $j$ particles coming from a source are in 
diagonal form in some basis $\lbrace\vert k\rangle\rbrace$, i.e.,
\begin{equation}
 \sum_{k=-j}^{k=j}p_{k}\vert k\rangle\langle k\vert=\sum_{k=1/2(0)}^{k=j}(p_{k}\vert +k\rangle\langle +k\vert + p_{-k} \vert -k\rangle\langle -k\vert) 
\end{equation}
where, $\sum_{k=-j}^{k=j} p_{k}=1$. 
We define an observable $Q$  following \citep{GP} in the way given below:
\begin{eqnarray}
Q=\frac{\Gamma_{z}  + \Pi}{\sqrt{2j+1}}\nonumber\\
=(\sigma^{1}_{z}\oplus\sigma^{2}_{z}+...\oplus\sigma^{j}_{z} + \Pi)/\sqrt{2j+1}
\end{eqnarray}
where, $\Pi$ is the null matrix when $N(=2j+1)$ is even, and for odd $N$ the
only nonvanishing element of $\Pi$ is $(\Pi)_{N,N}=\frac{1}{\sqrt{2}}$. In order to
maintain optimal violation of the four-term LGI,  we require our 
time-evolved observables to remain in the block diagonal form mentioned above. This can not be 
ensured by arbitrary rotations of the SG apparatus in space, except for two dimensional systems.  However, this is achieved if 
each block is evolved separately~\citep{GP}. As $\oplus_j\exp^{i\theta_{j}\sigma_{x}}\sigma_{z}\exp^{-i\theta_{j}\sigma_{x}}=\oplus_j\exp^{i\theta_{j}\sigma_{x}}\oplus_{j} \sigma_{z}^{j}\oplus_j\exp^{-i\theta_{j}\sigma_{x}}$, time 
evolution of the system is affected by
\begin{eqnarray}\label{uni}
U(t)=\exp^{-i\theta_{1}\sigma_{x}}\oplus \exp^{-i\theta_{2}\sigma_{x}}\oplus ...\oplus \exp^{-i\theta_{j}\sigma_{x}}
\end{eqnarray}
We explain in next paragraph explicitly how this kind of evolution and measurements are realised experimentally. First, the system is evolved to time $t_{1}$ and $Q$ is measured. The 
post-measurement state is further evolved to time $t_{2}$, and $Q$ is measured
again. This scheme can be recast into the Heisenberg picture. Taking all 
$\theta_{j}=\omega t/2=\alpha$, we have
\begin{eqnarray}
U^{\dagger}(t)Q U(t)=(\cos\alpha  \Gamma_{z}+\sin\alpha  \Gamma_{y}+ \Pi)/\sqrt{2j+1}.
\end{eqnarray}\\
 The two-time correlation function $C_{12}$ using lemma 1, for even $N$ is 
given by 
 \begin{eqnarray}\label{ce}
C_{12}=Tr[U^{\dagger}(t_{1}) Q U(t_{1})U^{\dagger}(t_{2}) Q U(t_{2})]\nonumber\\
=[ \cos\alpha_{1}\cos\alpha_{2}+\sin\alpha_{1}\sin\alpha_{2}].
\end{eqnarray}\\
For odd $N$, one gets
 \begin{eqnarray}\label{co}
C_{12}=\frac{2j( \cos\alpha_{1}\cos\alpha_{2}+\sin\alpha_{1}\sin\alpha_{2})+1/\sqrt{2}}{(2j+1)}.
\end{eqnarray}\\
Similarly, $C_{23}$ and $C_{34}$ are also obtained. For obtaining the
correlation function $ C_{14}$, 
the operator $Q$ is taken 
to be  $(\Gamma_{z}  - \Pi)/\sqrt{2j+1}$.  Now, in order to obtain the
maximal violation of the LGI, we choose the time intervals such that 
$\alpha_{1}=0,\alpha_{2}=\pi/4, \alpha_{3}=\pi/2, \alpha_{4}=3\pi/4$. Hence, the
 value of the Leggett-Garg sum for the spin $j$ system is given by $2\sqrt{2}$.
$~~~~~~~~~~~~~~~~~~~~~~~~~~~~~~~~~~~~~~~~~~~~\blacksquare$

 We now outline a conceptual scheme for verifying the correlations 
(\ref{ce},\ref{co}) following\citep{GP}. Such a  scheme has been experimentally implemented for 
spin-$1$ entangled particles~\citep{how}. For the present purpose, consider 
spin $j$ particles emerging from an initial ensemble. These particles are
assumed to possess not only a magnetic moment (an interaction energy 
$\mu B_{z}J_{z}$), but also an electric quadrupole moment (an interaction
energy proportional to $E_{z}J_{z}^{2}$). The particles first pass through 
inhomogeneous electric fields producing beams with 
$\vert m_{z}\vert=j,j-1...0(1/2)$. Taken separately, each of these beams 
with given $\vert m_{z}\vert$ are passed through a uniform magnetic field $B_{z}$ producing
energy difference, E, between them. Then an rf pulse generates Rabi oscillations of frequency
$\omega =E/\hslash$ among them. This captures the unitary evolution given by (\ref{uni}). After evolving for a time $t_{1}$, $\sigma_{z}$ is measured. Then, each post-measured beam is again evolved,  and at a time 
$t_{2}$, $\sigma_{z}$ is measured. This same procedure is done many times 
varying the time of measurements randomly. 
Correlation statistics are calculated from the measured data. For half-integral
 spin ($N$ even) this procedure is exact. For integral spin  (odd $N$), there 
is an unpaired beam corresponding to $m_{z}=0$ for which special treatment is 
needed to get maximum violation. The beam corresponding to $m_{z}=0$ is 
evolved by $U(t)=e^{-i\theta j_{y}}$, where, $j_{y}$ is the $y$-th component of 
the corresponding spin operator $j$. For example, for a spin $ 1$ system,
the evolution leads to
\begin{eqnarray}\label{evo0}
e^{-i\theta_{1} j_{y}}\vert m_{z}=0\rangle =\frac{\sin\theta_{1}}{\sqrt{2}}\vert -1\rangle +\cos\theta_{1}\vert 0\rangle -\frac{\sin\theta_{1}}{\sqrt{2}}\vert 1\rangle.
\end{eqnarray} 
Next,  measurement is performed by an inhomogeneous electric field,  which is 
a two-outcome measurement characterised by the projectors $P_{+}=P_{1}+P_{-1}$ 
and $P_{-}=P_{0}$. Here, $P_{\pm 1}$ is the projector on the subspace spanned by 
$m_{z}=\vert\pm 1\rangle$ components and $P_{0}$ is projector on the 
$\vert 0\rangle$ state. According to the outcome, the  post-measured states 
are $(\vert -1\rangle -\vert 1\rangle)/\sqrt{2}$ with probability $p_{+}=\sin^{2}\theta_{1}$, and $\vert 0\rangle$ with probability $p_{-}=\cos^{2}\theta_{1}$,
respectively. These post-measurement states are again evolved  to time $t_{2}$.
The  `$+$' outcome states evolve to $(\cos\theta_{2}\vert -1\rangle -\sqrt{2}\sin\theta_{2}\vert 0\rangle -\cos\theta_{2}\vert 1\rangle)/\sqrt{2}$,  and the
`$-$' outcome states evolve as given by Eq.(\ref{evo0}). The conditional 
probabilities are $p_{+|+}=\cos^{2}\theta_{2}, p_{-|+}=\sin^{2}\theta_{2}, p_{+|-}=\sin^{2}\theta_{2}, p_{-|-}=\cos^{2}\theta_{2}$, where $\theta_{2}=\omega\Delta t$. 
The two-time correlation function is $\cos (2\omega\Delta t)$, which is the 
same as in the qubit case, and hence gives the maximum violation up to 
$2\sqrt{2}$.

\section{\textbf{IV. Disappearance of violation through unsharp measurement}}

  The purpose of this section is to show that with sufficiently unsharp 
measurements, the outcome statistics can be described by classical theory.
Using a particular form of coarse graining it was shown in  Ref.~\citep{KB} 
that when the resolution of the apparatus is much greater than the intrinsic 
quantum uncertainty, i.e., $\Delta m >> \sqrt{j}$, the outcomes appear to obey 
classical laws. Under this condition the measurements become fuzzy enough for
the non-invasive assumption to become essentially valid and the system dynamics
 mimics the rotation of a classical spin coherent state. However, in this 
formalism there does not exist any sharp cut-off for the value of the apparatus
resolution beyond which classicality emerges. Or, in other words,
given a specific quantum system with a particular intrinsic uncertainty, it is
not clear as to what is the precise value of the apparatus resolution above
which the condition of coarse graining is satisfied.  

Here we follow the theory of unsharp observables~\citep{bookB} which as an 
element of unsharp reality provides the necessary ingredient in modelling of 
the emergence of classical behaviour within quantum mechanics in a precise and
quantitative manner. It has been  shown by Kar and 
Roy~\citep{GK} that for the value of the sharpness parameter 
$\lambda \leq 1/2^{1/4}$, the CHSH inequality is always satisfied for spin 
$1/2$ systems. As the spin or polarization observables of  entangled 
particles in an EPR experiment are measured with progressively more limited 
accuracy, there is a corresponding progressive degradation of the degree of 
Bell violation. Violation of the Bell inequality becomes unobservable above a 
certain degree of inaccuracy~\citep{BJ}.  
Unsharp measurement for a class of LGIs has been considered for two level 
systems in a recent work~\citep{wlgi} . Here we extended this approach for 
arbitrary spin systems. 

Let us first describe briefly the formalism of unsharp 
measurements~\citep{bookB} relevant to our present analysis.
In projector valued measurements the observables are self-adjoint operators 
having projectors as spectral, i.e., $A\equiv \lbrace P_{i}\vert\sum P_{i}=\mathbb{I},P_{i}^{2}=P_{i}\rbrace$. The probability of getting the $i$-th outcome is 
$tr[\rho P_{i}]$ for the state $\rho$. Extending to positive operator valued 
measures (POVM), the observables are self-adjoint operators but with spectral 
as positive operators within the interval $[0,\mathbb{I}]$, i.e., $E\equiv \lbrace E_{i}\vert\sum E_{i}=\mathbb{I},0< E\leq \mathbb{I}\rbrace$.  Similarly, the
 probability of getting the $i$-th outcome is $tr[\rho E_{i}]$. 
Effects ($E_{i}s$) represent quantum events that may occur as outcomes of a 
measurement. A subclass of effects of particular interest are the regular 
effects, characterized by the property that their spectrum extends both 
above and below the value $1/2$. For two outcome measurements this notion is 
captured by the effect, $E_{\lambda}=(\mathbb{I}+\lambda n_{i}\sigma_{i})/2, i=1,2,3.$, with $\lambda\in(0,1]$. Thus, the set of effects can be written as a 
linear combination of sharp projectors with white noise, $E_{\lambda}\equiv \lbrace E_{+}^{\lambda},E_{-}^{\lambda}\vert E_{+}^{\lambda}+E_{-}^{\lambda}=\mathbb{I}\rbrace$, given by
\begin{eqnarray}
E_{\pm}^{\lambda}=\frac{1+\lambda}{2}P_{\pm}+\frac{1-\lambda}{2}P_{\mp}\nonumber\\
=\lambda P_{\pm}+\frac{1-\lambda}{2}\mathbb{I}.
\end{eqnarray}\\
This can be thought of as projectors becoming noisy reflecting inaccuracy of 
the experiment. Hence, the  sharpness or dialation parameter $\lambda$ can be 
estimated from the difference between the really observable data and that 
predicted by sharp observables. Under this unsharp measurement, the state 
transformation for the maximally mixed initial state is given by the generalised L\"udders operation
 \begin{eqnarray}
 \rho_{\pm}^{PM}(t_{1})=\sqrt{ E_{\pm\lambda}}\rho\sqrt{ E_{\pm\lambda}}/tr[\sqrt{E_{\pm\lambda}}\rho\sqrt{ E_{\pm\lambda}}]\nonumber\\
 =\sqrt{\frac{1}{2}(\mathbb{I}\pm\lambda\sigma_{z})} \frac{\mathbb{I}}{2}\sqrt{\frac{1}{2}(\mathbb{I}\pm\lambda\sigma_{z})}\nonumber\\
 =\frac{1}{2}(\mathbb{I}\pm\lambda\sigma_{z}).
 \end{eqnarray}
 The probability of getting `$\pm$' outcomes are both $1/2$. In order to 
formulate the relevant LGI,  the $\rho_{\pm}^{PM}(t_{1})$ is evolved for time 
$\Delta t$, giving $ \exp^{-i\frac{\omega \Delta t}{2}\sigma_{x}}\frac{1}{2}(\mathbb{I}\pm\lambda\sigma_{z})\exp^{i\frac{\omega\Delta t}{2}\sigma_{x}}
 =\frac{\mathbb{I}}{2}\pm\frac{\lambda}{2}(\cos(\omega\Delta t)\sigma_{z}+\sin(\omega\Delta t)\sigma_{y})$.
 With this post-measurement state we find the conditional probabilities and 
the two-time correlation function given by $C_{12}=\lambda^{2}\cos(\omega\Delta t)$. Hence, the LGI with unsharp measurement can be written as  $K\equiv\lambda^{2}\langle LGI\rangle\leq 2$, where $\langle LGI\rangle$ denotes the 
corresponding expression for sharp measurements. Since $\langle LGI\rangle_{max}=2\sqrt{2}$, hence it follows that in the case of unsharp measurements the LGI 
for a spin $1/2$ system is always satisfied when the 
sharpness parameter upper bounded by $\lambda<1/2^{1/4}$.

 Now, let us consider a system having arbitrary spin. As discussed in our
conceptual scheme of measurement in the previous section,  particles of spin 
$j$  are sent from the source to an inhomogeneous electric field. After 
emerging from the field each beam is effectively described by a two dimensional
 Hilbert space, and evolves under the same unitary as above. Finally, the
beam is subjected to a non-ideal Stern-Gerlach apparatus~\citep{SG}. 
In this scenario 
the effective spin $j$ observable is given by 
$Q=\lambda((\Gamma_{z}+\Pi)/\sqrt{2j+1} )$, where $ 0<\lambda\leq 1$. Using 
the lemma in section III it is straightforward to calculate the two-time 
correlation function. When $N$ is even, we have
 \begin{eqnarray}
 C_{12}=\lambda^{2}[ \cos\alpha_{1}\cos\alpha_{2}+\sin\alpha_{1}\sin\alpha_{2}]
\end{eqnarray}
and for $N$ odd, it becomes
\begin{equation}
C_{12}=\lambda^{2}[2j \cos\alpha_{1}\cos\alpha_{2}+2j\sin\alpha_{1}\sin\alpha_{2}+1/\sqrt{2}]/(2j+1)
\end{equation}
For both the even and odd cases the optimal value of the sharpness parameter 
below which no quantum violation of LGI is possible is upper bounded by 
$1/2^{1/4} \equiv 0.841$. Note in comparison that using the maximal violation
of the LGI for large spin obtained in Ref.~\citep{KB}, the 
required value of this 
parameter would be $0.8978$ in order to ensure satisfaction of the LGI.
Note also, that for the case of spatial correlations, a higher value of the
sharpness parameter would be required to ensure satisfaction of the relevant
local realist inequality, since the maximal bound there drops for the case of
integral spin~\citep{GP}.  For the case of spin-$1$ particles the value
of the required sharpness for spatial correlations turns out to be $0.8852$ and 
coincides with our optimal value ($0.841$) for temporal correlations with
 infinitely large integral spin.

\section{\textbf{V. LG-CH inequalities and emergence of classicality}}

 We finally show how satisfaction of the LGI implies that there exists a 
non-invasive realist model for the temporal correlations as described in 
section II.  Following the line of reasoning used by Fine~\citep{fine} for the
case of spatial correlations, we show here that satisfaction of the LGI inequality is 
a necessary and sufficient condition for the existence of such a classical 
model in a situation when no-signaling in time (NSIT)~\citep{KB2} is satisfied. 
Fine's theorem~\citep{fine} showed that Bell-CH inequalities provide a necessary and 
sufficient condition for existence of joint probabilities of all measured observables
implying a local realist model.
Recently, a new proof of Fine's theorem has been presented~\citep{hal}. The similar nature of the local hidden variable models in the spatial 
scenarios and the non-invasive hidden variable models in the temporal 
scenario~\citep{fritz} enables the adoption of Fine's theorem  in 
the temporal domain.  To this end we first derive a CH type~\citep{CH} LGI, 
and make clear how satisfaction of such LGI-CH implies existence of a 
non-invasive realist model (NIRM). 

We assume the following properties of the joint and single probabilities, 
$p^{++}(Q_{1}Q_{2})+p^{+-}(Q_{1}Q_{2})=p^{+}(Q_{1})$, $p^{++}(Q_{1}Q_{2})+p^{-+}(Q_{1}Q_{2})=p^{+}(Q_{2}) $
and $p^{+}(Q_{1})+p^{-}(Q_{1})=1$, $p^{+}(Q_{2})+p^{-}(Q_{2})=1$. Now the two-time 
correlation function becomes $C_{12}=4 p^{++}(Q_{1},Q_{2})-2p^{+}(Q_{1})-2p^{+}(Q_{2})+1$. 
It is to be noted that obtaining single probabilities from joint probabilities amounts to 
satisfying the 
assumption of induction at the statistical level and no-signaling in time (NSIT). 
Unlike in the Bell scenario where the no-signaling principle holds, temporal correlation can 
violate NSIT. However,
induction is always satisfied.
It is shown in Ref.~\citep{KB2} that NSIT is an alternative necessary condition of MR. 

The normalization condition for the single probabilities implies that 
ideal lossless detectors are used for measuring outcomes. Writing the two-time 
correlation in this form one can derive an inequality in a form equivalent
to the Bell-CH inequality~\citep{CH} in the temporal domain. We call such an
inequality as an LGI-CH inequality, given by 
\begin{eqnarray}
-1\leq p^{++}(Q_{1}Q_{2})+p^{++}(Q_{3}Q_{2})-p^{++}(Q_{1}Q_{4})\\ \nonumber+p^{++}(Q_{3}Q_{4}) -p^{+}(Q_{3})-p^{+}(Q_{2})\leq 0.
\end{eqnarray}
Under the above conditions other LGI-CH inequalities can be derived by varying 
outcome combinations. These LGI-CHs may be combined to obtain the LGI. Now, it 
is straightforward following the line of resoning presented by 
Fine~\citep{fine} that satisfaction of these inequalities imply the existence
of joint probability distributions for all the observables, i.e., $p(Q_{1},Q_{2},Q_{3},Q_{4})$.  This in turn implies a noninvasive realist model
that mimics the temporal correlations for the measurement of a dichotomic observable at time $t_{1}, t_{2}, t_{3}, t_{4}$.
Such a model is compatible with classical theory, and the quantum dynamics of 
the system boils down to some classical stochastic process with the 
measurement statistics  given by averaging over this process. In precise we prove that
\begin{eqnarray}
LGI \wedge NSIT <==> LG-CH <==> NIRM 
\end{eqnarray}\\

\section{\textbf{VI.  Conclusions}}

To summarize, in the present work we have shown that  optimal violation of 
the Leggett-Garg inequality~\citep{lgi1,lgi2} is allowed by 
quantum theory in the context of a suitably adopted measurement scheme for 
a system possessing arbitrary spin. The observable~\citep{GP} chosen 
here enables one to
achieve the maximal limit of temporal correlations irrespective of the integral
or half-integral value of the spin, improving upon earlier results obtained 
through the choice of other observables~\citep{KB}.  It may be noted that
whereas we obtain maximal violation of macrorealism for an arbitrary spin 
system, the same
 Peres-Gisin observable~\citep{GP} used in the case of spatial correlations
does not lead to maximal violation of the corresponding local realist 
inequality for finite integral spin systems.  

We have further shown how 
 coarse graining of the measurement process through unsharp 
observables~\citep{bookB} leads 
to the satisfaction of LGI. The form of coarse graining used here is 
quantitative, as different from the coarse-graining employed in a similar
context earlier~\citep{KB}.  Here it is possible to obtain the  precise 
threshold value of the sharpness parameter below which no quantum violation of 
the LGI can be achieved.  Our approach using unsharp measurements fits 
naturally within the context of non-ideal apparatus~\citep{SG} in actual 
experimental conditions. Comparing with the case of spatial correlations
for a similar coarse grained approach through unsharp measurement, we find
here that for temporal correlations the satisfaction of LGI
emerges below a smaller value of the sharpness parameter.  This lends 
further credence to the
contention~\citep{fritz} of temporal correlations being somehow `stronger' 
than spatial correlations. We finally show that satisfaction of the the LGI
implies existence of a non-invasive realist model for dichotomic measurement 
at four different times in a situation when NSIT is satisfied.

Note : Recently, Clemente and Kofler have shown in Ref.\citep{ck, ck2} how various (more than 
two time) NSIT and Arrow-of-Time (induction) form a necessary and sufficient condition for 
MR~\citep{we}.

{\it\textbf{Acknowledgement:}} The authors acknowledge  useful 
discussions with D. Home and G. Kar. We thank A. Fine for helpful clarifications on Fine's 
theorem and 
making us aware of Ref.~\citep{hal}. We also thank A. Brodutch,  L. Clemente and J. Kofler for 
useful comments. ASM acknowledges support from the project SR/S2/LOP-08/2013 of DST,
India.


\begin{thebibliography}{99}

\bibitem {Bell} J. S. Bell, Physics (N.Y.) 1, 195 (1965); J. S. Bell, 
\emph{Speakable and Unspeakable in Quantum Mechanics} (Cambridge University Press, Cambridge, England, 2004).

\bibitem{CHSH} J. F. Clauser, M. A. Horne, A. Shimony and R. A. Holt, Phys. Rev. Lett. {\bf 23}, 880 (1969).

\bibitem{tsi} B. S. Cirel'son, Lett. Math. Phys. {\bf4}, 93 (1980).

\bibitem{vanden} L. Vandenberghe and S. Boyd, SIAM Rev. {\bf 38}(1), 49 (1996).

\bibitem{wehner} S. Wehner, Phys. Rev. A {\bf 73}, 022110 (2006); M. Navascues,
 S. Pironio, and A. Acin., Phys. Rev. Lett. {\bf 98}, 010401 (2007).

\bibitem{mer1} N. D. Mermin, Phys. Rev. Lett.  {\bf 65}, 1838 (1990);
S. M. Roy and V. Singh, Phys. Rev. Lett. {\bf 67}, 2761 (1991).

\bibitem{M}A. Garg, N. D. Mermin, Phys. Rev. Lett. {\bf 49}, 1220
(1982); A. Garg, N. D. Mermin, Phys. Rev. D {\bf 27}, 339 (1983).

\bibitem{P} A. Peres, Phys. Rev. A {\bf 46}, 4413 (1992). 
\bibitem{GP} N. Gisin, A. Peres, Phys. Lett. A {\bf 162}, 15 (1992).


\bibitem{HM} D. Home, A. S. Majumdar, Phys. Rev. A {\bf 52}, 4959 (1995);
A. Cabello, Phys. Rev. A {\bf 65}, 062105 (2002).

\bibitem{GM} A. Garg and N. D. Mermin, Phys. Rev. Lett. {\bf 49}, 901 (1982);
N. D. Mermin, Phys. Rev. Lett. {\bf 65}, 3373 (1990);
N. D. Mermin, Rev. Mod. Phys. {\bf 65}, 803 (1993); M. Zukowski and D.
Kaszilokowski, Phys. Rev. A {\bf 56}, R1682 (1997); R. F. Werner and M. M.
Wolf, Phys. Rev. A {\bf 64}, 032112 (2001). 

\bibitem{how}
J. C. Howell, A. Lamas-Linares, 
and D. Bouwmeester, Phy. Rev. Lett. {\bf 88} 030401 (2002).

\bibitem{lgi1} A. J. Leggett and A. Garg, Phys. Rev. Lett. {\bf 54}, 857 (1985).

\bibitem{lgi2} A. J. Leggett, J. Phys. Condens. Matter {\bf 14}, R415 (2002).

\bibitem{KB} J. Kofler, C. Brukner Phys. Rev. Lett. {\bf 99}, 180403 (2007).

\bibitem{avis} D. Avis, P. Hayden, and M. M. Wilde, Phys. Rev. A {\bf 82}, 030102(R) 
(2010); A. R. U. Devi, H. S. Karthik, Sudha, and A. K. Rajagopal, Phys. Rev. 
A {\bf 87}, 052103 (2013); 
D. Gangopadhyay, D. Home, and A. S. Roy, Phys. Rev.
A {\bf 88}, 022115 (2013);
S. Das, S. Aravinda , R. Srikanth, D. Home, Europhys. Lett. {\bf 104},  
60006 (2013); S. Mal, M. Banik, S. K. Chaudhary,  arXiv:1406.5754 (2014).
\bibitem{KB2}  J. Kofler and C. Brukner, Phys. Rev. A {\bf 87}, 052115 (2013);

\bibitem{budroni} C. Budroni, T. Moroder, M. Kleinmann, and O. Guhne; Phys. 
Rev. Lett. {\bf 111}, 020403 (2013);  C. Budroni and C. Emary, Phys. Rev. 
Lett. {\bf 113}, 050401 (2014).

\bibitem{revw} A. J. Leggett, Rep. Prog. Phys. {\bf 71}, 022001 (2008); 
C. Emary, N. Lambert, and F. Nori, Rep. Prog. Phys. {\bf 77}, 016001 (2014).

\bibitem{ex1} C. H. van der Wal et al., Science {\bf 290}, 773 (2000); J. R. 
Friedman et al., Nature {\bf 406}, 43 (2000); R. Roskov, A. M. Korotkov, and 
A. Mizel, Phys. Rev.
Lett. {\bf 96}, 200404 (2006); A. N. Jordan, A. M. Korotkov,
and M. Buttiker, Phys. Rev. Lett. {\bf 97}, 026805 (2006); 
V. Athalye, S. S. Roy, and T. S. Mahesh, Phys. Rev. Lett.
{\bf 107}, 130402 (2011).

\bibitem{onof} T. Calarco and R. Onofrio, Phys. Lett. A {\bf 198}, 279 (1995);
{\it ibid.} {\bf 208}, 40 (1995); T. Calarco, M. Cini and R. Onofrio,
Europhys. Lett. {\bf 47}, 407 (1999).

\bibitem{peres1} A. Peres, Phys. Lett. A {\bf 151}, 107 (1990); 
A. Peres, {\it Quantum Theory: Concepts and Methods},
(Kluwer, Dordrecht, 1993).

\bibitem{fritz} T. Fritz, New Journal of Physics {\bf 12}, 083055 (2010).

\bibitem{hardy} L. Hardy, Phys. Rev. Lett. {\bf 71}, 1665 (1993).

\bibitem{bookB} P. Busch, Phys. Rev. D {\bf 33} 2253 (1986); P. Busch, M. 
Grabowski, and P. J. Lahti, {\it Operational Quantum
Physics, Vol. 31}, (Springer, Berlin, 1995).

\bibitem{BJ}
P. Busch, G. Jaeger, Found. Phys.{\bf 40}, 1341 (2010).

\bibitem{fine} A. Fine, Phys. Rev. Lett. {\bf 48}, 291 (1982).

\bibitem{wlgi} D. Saha, S. Mal, P. K. Panigrahi, D. Home, Phys.Rev. A {\bf 91}, 
032117 (2015).

\bibitem{years} O. J. E.  Maroney, arXiv: 1207.3114 (2012), J. M. Yearsley,  
arXiv: quant-ph 1310.2149 (2013).

\bibitem{GK} G. Kar, S. Roy, Phys. Lett. A {\bf 199}, 12 (1995).

\bibitem{SG} D. Home, A. K. Pan, Md. M. Ali, A. S. Majumdar, J. Phys.
A: Math. Theor. {\bf 40}, 13975 (2007); B. C. Hsu, M. Berrondo and 
J.-F. S. Van Huele, Phys. Rev. A {\bf 83}, 012109 (2011); A. K. Pan and A. 
Matzkin, Phys. Rev.
A {\bf 85}, 022122 (2012).

\bibitem{CH} J. F. Clauser, M. A. Horne, Phys. Rev. D {\bf 10}, 526 (1974).
\bibitem{hal}J. J. Halliewll, Phys. Lett. A {\bf 378}, 2945 (2014).
\bibitem{ck}L. Clemente and J. Kofler, Phys. Rev. A {\bf 91}, 062103 (2015).
\bibitem{ck2}L. Clemente and J. Kofler, arXiv: 1509.00348 (2015).

\bibitem{we} Fine's theorem allows us to show that satisfaction of LGIs and the two time NSIT 
conditions are necessary and sufficient for MR involving 
dichotomic measurements at four different times, which is the case dealt here.












\end{thebibliography}
\end{document}